\documentstyle[prl,aps]{revtex}
\input{epsf}

\def\be{\begin{equation}}
\def\ee{\end{equation}}
\def\bea{\begin{eqnarray}}
\def\eea{\end{eqnarray}}
\def\half{\frac{1}{2}}
\def\sixth{\frac{1}{6}}

\begin{document}

\title{\bf A two-species continuum model for aeolian sand ripples}

\author{Rebecca B. Hoyle}
\address{Department of Applied Mathematics and Theoretical Physics,
Silver Street, Cambridge CB3 9EW, UK. \\
email: R.B.Hoyle@damtp.cam.ac.uk}
\author{Anita Mehta }
\address{S.N. Bose National Centre for Basic Sciences, Block JD,
Sector III Salt Lake,
Calcutta 700 091, India. \\
email: anita@boson.bose.res.in}
\date{\today}
\maketitle

\begin{abstract}
We formulate a continuum model for aeolian sand ripples consisting
of two species of grains: a lower layer of relatively immobile clusters,
with an upper layer of highly mobile grains moving on top.
We predict analytically the ripple wavelength, initial ripple growth rate
and threshold saltation flux for ripple formation.
Numerical simulations show the evolution of realistic ripple profiles from
initial surface roughness via ripple growth and merger.
\end{abstract}


\narrowtext

Aeolian sand ripples are formed by the action of the wind on the sand
bed in the desert and at the seashore. They have also recently been observed 
on Mars~\cite{mars}. Aeolian ripples are a few centimetres in 
wavelength and their crests lie perpendicular to the prevailing wind 
direction.  
Bagnold (1941) made an influential early study~\cite{bagnold}. He identified
the importance of saltation, where sand grains are entrained
by the wind, and whipped along the sand bed, colliding with it at
 high speed (of order $\le$ 
1m/s~\cite{andersbun}), and
causing other grains to jump out of and along the bed, thus sculpting ripples. 
The impact angle remains roughly constant at about $10^\circ - 16^\circ $ 
despite the gusting of the wind. The stoss slope of the ripple lies in
the range  $8^\circ$-$10^\circ$~\cite{sharp}. The lee slope is composed of a 
short straight section
near the crest at an angle of about $30^\circ-34^\circ$ to the 
horizontal~\cite{bagnold}~\cite{sharp}~\cite{pye}, 
followed by a longer and shallower concave section. The deposition of grains 
on the lee slope of sand dunes leads to 
oversteepening and avalanching~\cite{hunt}~\cite{anders3}, which
maintains the lee slope at an angle of around $30^\circ-34^\circ$
near the crest. A similar, if less dramatic, mechanism is likely to hold
for ripples.

Numerical simulations of ripples and dunes based on tracking individual sand 
grains or on cellular automata have in the past few years yielded good 
qualitative agreement with observations~\cite{anders}. 
However, these 
methods are computationally expensive. Continuum models provide
a complementary approach allowing
 faster calculation of ripple evolution, and the possibility of obtaining 
certain information, such as the preferred ripple wavelength, analytically.
Anderson~\cite{anders2} produced an analytical model of the initial 
generation of ripples from a flat bed. A one-species analytical
continuum model was formulated by Hoyle and Woods~\cite{menandy}.
In the current Letter we extend the one-species model~\cite{menandy} to 
include relaxation effects, inspired by models of sandpiles~\cite{anita1}, 
in order to obtain 
more realistic ripple profiles and to predict the ripple wavelength and speed.
As our work was nearing completion we learned of other work in 
progress on
ripple formation~\cite{prig} similar in spirit to our own, but with 
important differences in implementation.

We consider two-dimensional sand ripples: following earlier work on
sandpiles~ \cite{anita1}~\cite{anita0} we  
assume that the effective surface of the ripple
comprises a `bare' surface defined by the local heights of
clusters $h(x,t)$ sheathed by a thin layer of flowing
mobile grains whose local density is
 $\rho(x,t)$, with 
$x$ the horizontal coordinate, and $t$ time.
The ripple evolves under the influence of two distinct types of mechanism.
Firstly, the impact of a constant flux of saltating
grains, knocks grains out of the `bare' surface, causing them to
hop along the ripple surface and land in the layer of flowing grains. 
This is the underlying
cause of ripple formation. Secondly, the ripples are subject to
intracluster and intercluster granular relaxation mechanisms which result
in a smoothing of the surface.

Following~\cite{menandy} we consider grains to be bounced out of the `bare'
surface by a constant incoming saltation flux, which impacts the sand 
bed at an angle $\beta$ to the horizontal.
These hopping or `reptating'~\cite{mith} 
grains subsequently land in the flowing layer. The saltating grains 
are highly
 energetic and continue in saltation upon rebounding from the sand bed.
We assume that the number $N(x,t)$
of sand grains ejected per unit time, per unit length of the sand bed, is 
proportional to the component of the saltation flux perpendicular to the 
sand surface, giving
\be
N(x,t) = J \sin (\alpha + \beta) = 
{J (\sin \beta + h_x \cos \beta) \over (1+h_x^2)^{1/2}},  
\ee
where $\alpha = \tan^{-1}(h_x)$, and $J$ is a positive
constant of proportionality.
We assume that each sand grain ejected from the surface hops a 
horizontal distance $a$, with probability $p(a)$,
and then lands in the flowing layer. We consider the flight of each sand grain
 to take place instantaneously, since ripples evolve on a much slower 
timescale than that of a hop. The hop length distribution $p(a)$ can be 
measured experimentally~\cite{mith}~\cite{ung}, so we consider it
to be given empirically. It is possible that $p(a)$ and hence the mean and 
variance of hop lengths, could depend upon factors such as wind speed.
The number $\delta n_o(x,t)$ of sand grains leaving the surface between
positions $x$ and $x + \delta x$ in time $\delta t$, where $\delta x$ 
and $\delta t$ are infinitesimal is given by
$\delta n_o(x,t) = N(x,t) \delta x \delta t$.
The change $\delta h$ in the surface height satisfies 
$\delta x \delta h(x,t) = -a_p \delta n_o(x,t)$,
where $a_p$ is the average cross-sectional area of a sand grain.
In the limit $\delta t \rightarrow 0$ we find that the contribution
to the evolution equation
for $h(x,t)$ from hopping alone is
\be 
h_t = -a_p N(x,t) =  -a_p J 
{ \sin \beta + h_x \cos \beta \over (1+h_x^2)^{1/2}}. 
\ee
There may be regions on the sand bed which are shielded from the incoming
saltation flux by higher relief upwind. In these regions there will be
no grains bounced out of the surface, and there will be no contribution
to the $h_t$ equation from hopping.
The number $\delta n_i(x,t)$ of sand grains arriving on the layer of
flowing grains between positions $x$ and $x + \delta x$ in time $\delta t$
is given by 
\be
\delta n_i(x,t) = \int_{-\infty}^{+\infty} p(a) N(x-a,t) da \delta x \delta t.
\ee
The change in depth of the flowing layer satisfies
$\delta x \delta \rho (x,t) = a_p \delta n_i (x,t)$,
and hence the contribution to the evolution equation for the flowing layer 
depth from
hopping alone is
\be
\rho_t = a_p J \int_{-\infty}^{+\infty} p(a) { \sin \beta + h_x(x-a,t)
 \cos \beta \over (1+h_x^2(x-a,t))^{1/2}} da. 
\ee

We incorporate diffusive motion~\cite{sam} as well as processes
governing the transfer between flowing grains and clusters, 
following~\cite{anita1}, leading to the equations
\bea
h_t &&= D_h h_{xx} -T(x,t) -a_p J 
{ \sin \beta + h_x \cos \beta \over (1+h_x^2)^{1/2}},  \\
\rho_t &&= D_\rho \rho_{xx} +\chi (\rho h_x)_x + T(x,t)  \nonumber \\
&&+ a_p J \int_{-\infty}^{+\infty} p(a) { \sin \beta + h_x(x-a,t)
 \cos \beta \over (1+h_x^2(x-a,t))^{1/2}} da,  
\eea
where $D_h$, $D_\rho$ and $\chi$ are positive constants and 
where $T(x,t)$, which represents the transfer terms, is given by
\be
T(x,t)= - \kappa \rho h_{xx} + \lambda \rho (|h_x| - \tan \alpha), 
\ee
for $0 \le |h_x| \le \tan \alpha$ and by
\be 
T(x,t)= - \kappa \rho h_{xx}  + {\nu (|h_x| - \tan \alpha) \over
(\tan^2 \gamma - h_x^2)^{1/2}},
\ee 
for $ \tan \alpha \le |h_x| < \tan \gamma $,
with $\kappa$, $\lambda$ and $\nu$ also positive constants.
The term $D_h h_{xx}$ represents the diffusive rearrangement of clusters
while the term $D_\rho \rho_{xx}$ represents
the diffusion of the flowing grains.
The flux-divergence term $\chi (\rho h_x)_x$ models the flow of surface
grains under gravity. The current of grains is assumed proportional to the
number of flowing grains and to their velocity, which in turn is proportional
to the local slope to leading order~\cite{menandy}.
 The $- \kappa \rho h_{xx}$ term represents the inertial filling in of dips and
knocking out of bumps on the `bare' surface caused by rolling grains flowing
over the top.
The $\lambda \rho (|h_x| - \tan \alpha)$ term represents the tendency of 
flowing grains to stick onto the ripple surface at small slopes; it
 is meant to model the accumulation of slowly flowing grains
at an obstacle.  Clearly
for this to happen the obstacle must be stable, or else it
would be knocked off by the oncoming grains, so that this term only 
comes into play for slopes less than $\tan \alpha$, where $\alpha$ is
the  {\em angle of repose}.
The term $\nu(|h_x| - \tan \alpha) (\tan^2 \gamma - h_x^2)^{-1/2}$ represents
tilt and avalanching; it comes into play only for slopes greater than $\tan
\alpha$ and models the tendency of erstwhile stable clusters to shed grains
into the flowing layer when tilted.  This shedding of grains starts when the
surface slope exceeds the angle of repose. 
For slopes approaching the angle $\gamma$, which is the {\em maximum angle  
 of stability}, the rate of tilting out of grains becomes very
large: an avalanche occurs. This term, among other things, is a novel
representation of the well-known phenomena
of {\em bistability} and {\em avalanching} at the angle of repose~\cite{nagel}.

We renormalise the model equations setting
$x \rightarrow x_0 \tilde x$, $t \rightarrow t_0 \tilde t$,
$a \rightarrow x_0 \tilde a$, $\rho \rightarrow \rho_0 \tilde \rho$, 
$h \rightarrow h_0 \tilde h$, where
$x_0 = D_h/a_pJ \cos \beta$, $ t_0 = D_h /(a_pJ \cos \beta)^2$,
$h_0 = D_h \tan \gamma /a_pJ \cos \beta$, $\rho_0 = a_pJ \sin \beta/
\lambda \tan \alpha$,
which gives for 
$0 \le h_x \le \tan \alpha / \tan \gamma$
\bea
 h_t &&=(1+ \hat \kappa \rho) h_{xx} - \rho {\tan \beta \over \tan \alpha}
 \left ( |h_x| - {\tan \alpha \over  \tan \gamma} \right ) - f(x), \label{mod1}
\\
 \rho_t &&={h_0 \over \rho_0} \left \{ -\hat \kappa 
\rho h_{xx} + \rho {\tan \beta \over \tan \alpha} \left( |h_x| - {\tan \alpha
 \over \tan \gamma}\right ) \right \} \nonumber \\
&&+ {h_0 \over \rho_0}\int_{-\infty}^{+\infty}p(a)f(x-a)da
+ {D_\rho \over D_h} \rho_{xx}+\hat \chi (\rho h_x)_x \label{mod2} , 
\eea
 and for  $\tan \alpha /\tan \gamma \le h_x < 1$ 
\bea
 h_t &&=(1+ \hat \kappa \rho) h_{xx} - f(x) -{ \hat \nu  (|h_x|
- {\tan \alpha /
\tan \gamma} ) \over (1-h_x^2)^{1/2}}, \label{mod3} \\
 \rho_t &&={h_0 \over \rho_0}
\left \{-\hat \kappa \rho h_{xx}
+ { \hat \nu  (|h_x|- {\tan \alpha /
\tan \gamma}) \over (1-h_x^2)^{1/2}}\right \} \nonumber \\
&&+ {h_0 \over \rho_0}\int_{-\infty}^{+\infty}p(a)f(x-a)da
+{D_\rho \over  D_h} \rho_{xx} +\hat \chi (\rho h_x)_x, \label{mod4}
\eea 
where the tildes have been dropped and where
\be
f(x) =\left (h_x + {\tan \beta \over \tan \gamma}\right )(1+h_x^2 \tan^2
\gamma)^{-1/2},
\ee and
$\kappa = \kappa \rho_0/D_h$,
$\hat \nu = \nu t_0/h_0$,
$\hat \chi = \chi h_0/D_h$. 
Wherever the sand bed is shielded from the saltation flux, the hopping term 
must be suppressed by 
removing the term $-f(x)$ in the $h_t$ equation.  

Close to onset of the instability that gives rise to sand ripples, the slopes 
of the sand bed will be small, since surface roughness is of small
amplitude; hence the regime $0 \le \tan \alpha / \tan \gamma$
is relevant. There are no shielded regions at early times, since
the slope of the bed does not exceed $\tan \beta$. 
Note that $h_x=0$, $\rho=1$ is a stationary solution of equations (~\ref{mod1})
and (~\ref{mod2}).
Setting $h=\hat h e^{\sigma t+ikx}$ 
and $\rho=1+\hat \rho e^{\sigma t+ikx}$, where $\hat h \ll 1$ and 
$\hat \rho \ll 1$ are 
constants, linearising, and Taylor-expanding the integrand gives a dispersion
relation for $\sigma$ in terms of $k$.
The presence of the $|h_x|$ term means that
strictly we are considering different solutions for 
sections of the ripple where $h_x>0$ and sections where $h_x<0$, but in fact
the effect 
of the $|h_x|$ terms appears only as a contribution 
$\pm \tan \beta / \tan \alpha$ to the 
bracket $(1 \pm \tan \beta / \tan \alpha)$, where the $h_x>0$ case takes the 
$+$ sign and $h_x<0$ case takes the $-$ sign. Since typically we have
$\tan \beta /\tan \alpha \ll 1$, the two solutions will not be very different. 
One growth rate eigenvalue is given by 
$\sigma = - h_0 \tan \beta /\rho_0 \tan \gamma +O(k)$
and is the rate of relaxation of $\rho$ to its equilibrium value of $1$.
To $O(k^4)$ the other eigenvalue is
\be
\sigma = (\bar a -1 - \hat \chi \rho_0/h_0)k^2 + iAk^3 +Bk^4
\ee
where 
\bea
A &=& -\half \overline{a^2} + {\rho_0 \over h_0}{\tan \gamma \over \tan \beta}
\left ( 1 + \hat \chi {\rho_0 \over h_0} - \bar a -{D_\rho \over D_h} \right ) 
\left ( 1 \pm {\tan \beta
\over \tan \alpha} \right ),  \\
B &=& -\sixth \overline{a^3} - \half \overline{a^2}{\rho_0 \over h_0}{\tan 
\gamma 
\over \tan \beta}\left ( 1 \pm {\tan \beta \over \tan \alpha} \right ) 
-{\rho_0 \over h_0}{\tan \gamma \over \tan \beta}
\left (\bar a -1 -\hat \chi {\rho_0 \over h_0} +{D_\rho \over D_h} \right ) 
\left \{ \bar a + \hat \kappa + {\rho_0 \tan \gamma 
\over h_0 \tan \beta} \left ( 1 \pm {\tan \beta \over \tan \alpha} \right )^2 
\right \}, 
\eea
and where $\overline{(.)}$ denotes $\int_{-\infty}^{+\infty} (.) p(a) da$.
We have neglected higher order terms as we are looking for 
 long wave modes where $|k|$ is small, since short waves 
are damped by the diffusion terms. Sand ripples grow if $\bar a 
> 1+\hat \chi \rho_0/h_0$, 
which is equivalent to requiring that
$\bar a a_pJ \cos \beta > D_h + \chi a_p J \sin \beta / \lambda \tan \alpha$
  holds in physical variables, 
giving a threshold
saltation flux intensity for ripple growth. This is in agreement with the 
threshold found in~\cite{menandy}. Since $B$ is negative ($\beta < \alpha$), 
the fastest growing
mode has  wavenumber  $k^2 = -(\bar a -1-\hat \chi \rho_0/h_0)/2B$ 
with growth rate 
$\sigma = -(\bar a -1-\hat \chi \rho_0/h_0)^2/4B$. The allowed
band of wavenumbers for growing modes 
is $0< k^2 < -(\bar a -1-\hat \chi \rho_0/h_0)/B$.
The wave speed is given by $c = - A k^2 >0$; it is
higher for larger $k^2$ which implies that shorter waves  
move faster, as indeed was seen in the numerical simulations described below. 
The speed is higher for $h_x>0$ than
for $h_x<0$, leading to wave steepening.

The renormalised 
model equations (~\ref{mod1})-(~\ref{mod4}) were integrated numerically
using compact finite differences~\cite{lele} with periodic
boundary conditions.
The $-f(x)$ term in the 
$h_t$ equation was suppressed in shielded regions.
We used a normal distribution for the hop lengths with mean $\bar a$ and 
variance $s^2$.
In the run illustrated, we chose $\bar a = 3.1$, $s=0.1$, 
$D_\rho/D_h=1.0$,
$h_0/\rho_0=20.0$, $\hat \chi = 0.1$, $\hat \nu = 1.0$,
$\hat \kappa=0.1$, $\beta=10^\circ $, $\alpha=30^\circ $ and
 $\gamma=35^\circ$. The angles were chosen to agree with observational 
evidence, the ratio $h_0/\rho_0$ to ensure a thin layer of flowing grains,
and the remaining parameters to allow ripple growth.
The output was rescaled back into physical variables
using $D_h=1.0$ and $\lambda=10.0$. The initial conditions for the 
dimensionless variables were $h=1.0 + 0.1 \eta_h$, $\rho =0.95+0.1 \eta_\rho $,
where $\eta_h$ and $\eta_\rho$ represent random noise generated by
 random variables on $[0,1)$ in order to model surface roughness. In this
case $B=-8.47$ and $A=-5.26$ (taking the minus sign in the brackets), giving a 
preferred wavenumber of 
$k=0.352$, and a wave speed of $c=5.26k^2$. The length of
the integration domain was chosen to be ten times the linearly 
preferred wavelength.

\begin{figure}
\centerline{\epsffile{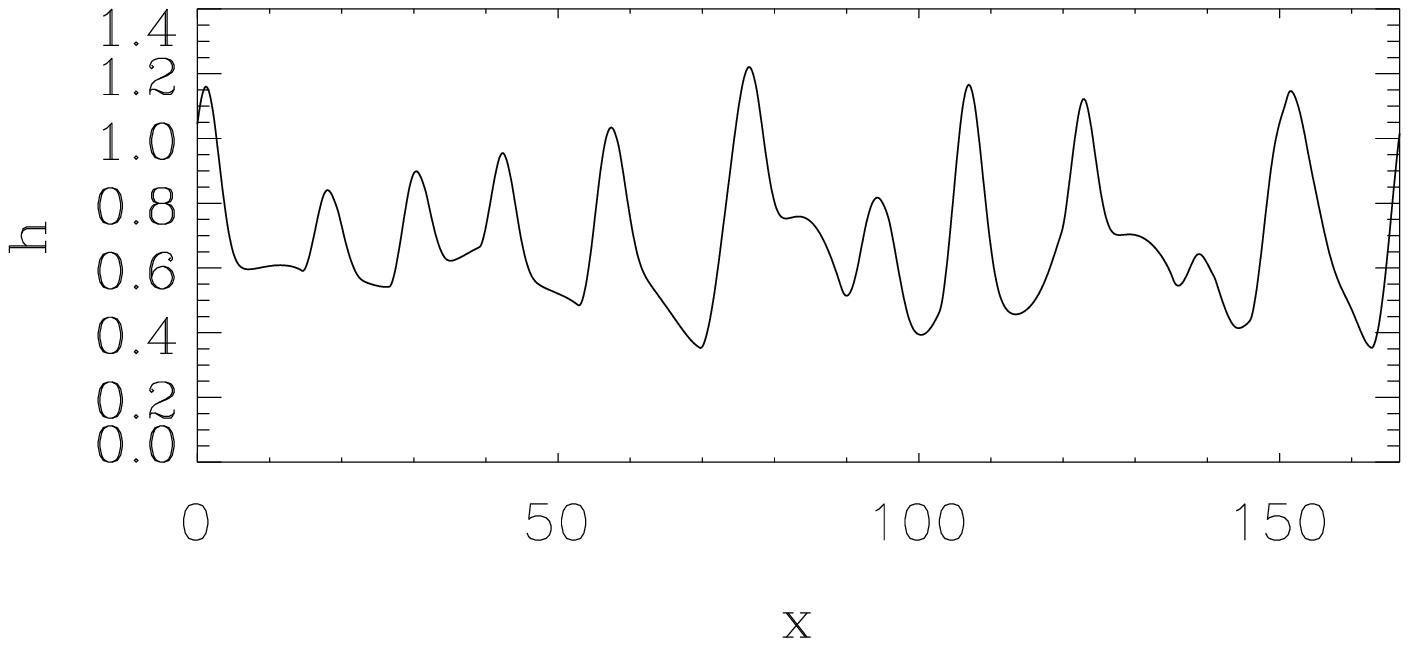}}
\caption{\it The surface height $h$ at
time $t= 10.0 \Delta t$.}
\label{fig01}
\end{figure}

\begin{figure}
\centerline{\epsffile{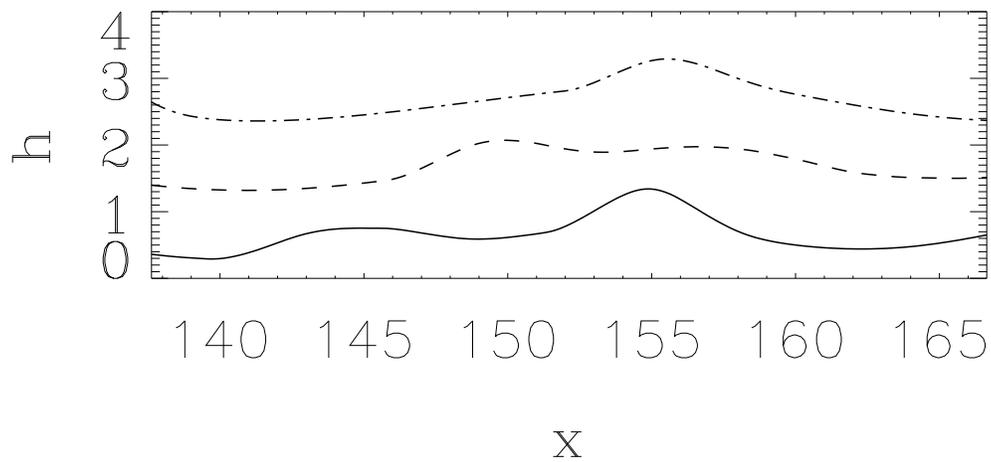}}
\caption{\it A sequence of profiles showing a small
ripple catching up and merging with a larger ripple. Sand is transferred
from the larger to the smaller ripple until only the latter remains. The 
profiles are
shown at times $t=11.0 \Delta t$ (solid line),
$t=12.0 \Delta t$ (dashed line) and $t=13.0 \Delta t$ 
(dot-dash line). The later profiles are each offset by one additional
unit in height.}
\label{fig03}
\end{figure}

\begin{figure}
\centerline{\epsffile{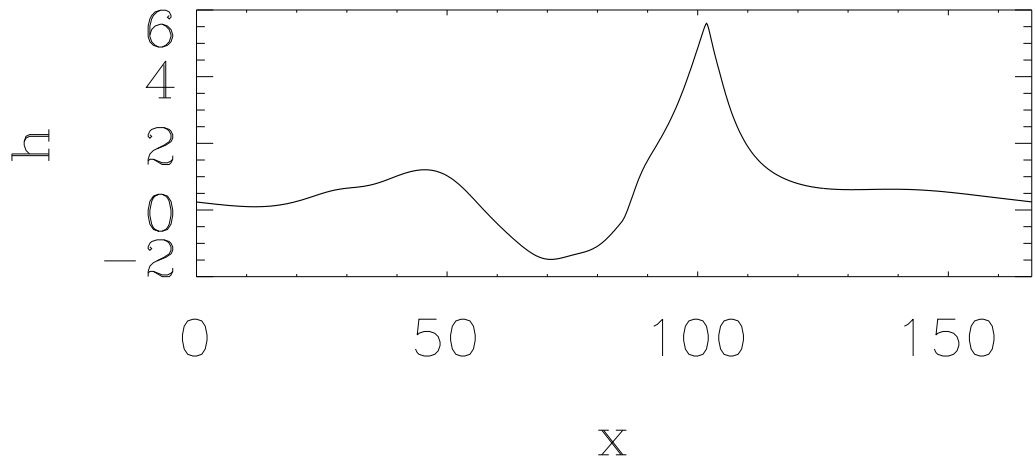}}
\caption{\it
 The surface height $h$ at time $t=89 \Delta t$, showing fully developed
ripples. Note the 
straight segments on the lee slopes close to the crests.}
\label{fig02}
\end{figure}

Figure~\ref{fig01}
 shows the surface height
at time $t= 10.0 \Delta t$, where $\Delta t=2.78$.
 Note the emergence of a 
preferred wavelength, with wavenumber $k \approx 0.457$ 
lying in the permitted band for
 growing modes predicted by the linear stability analysis
 and arrived at by a process of ripple merger. The wave speed close to
onset was also measured
and found to be $c \approx (4.13 \pm 0.39)k^2$, which is reasonably close 
to the 
predicted value. Ripple merger typically
occurs when a small fast ripple catches up and merges with a larger slower
ripple (figure~\ref{fig03}), the
leading ripple transferring sand to its pursuer until only the pursuer
remains. Occasionally a small ripple emerges from the front of the new
merged ripple and runs off ahead.
Figure~\ref{fig02} shows the surface height $h$ at 
 time $t=89 \Delta t$, with one shallow and one fully developed ripple.
Note the long shallow
stoss slopes, and the shorter 
steeper lee slopes with straight sections 
near the crests and concave tails. 
The leftmost ripple has a maximum stoss slope angle of
$3.3^\circ$, and a maximum lee slope of $9.2^\circ$, whereas the more fully
developed rightmost ripple has a maximum stoss slope angle of $24.8^\circ$
and a maximum lee slope angle of $33.5^\circ$, which lies between the
angle of repose and the maximum angle of stability.
The height to length ratio of the ripples is in the range
1:8 - 1:22,
which is in reasonable agreement with observations~\cite{sharp}.
In the long time limit, we would expect sand ripples to grow until the
maximum lee slope angle reaches an angle close to $\tan \gamma$. In
reality, there is only a relatively shallow layer of loose sand available
for incorportion into ripples, and this together with the maximum slope
condition will determine the size of the fully-developed ripples. 

In summary, we have formulated an analytical continuum model for aeolian 
sand ripples using a two-species model embodying intracluster and intercluster
relaxation, 
in a description that leads naturally to
bistable behaviour at the angle of repose,
and its cutoff at the angle of maximal stability~\cite{nagel}.
 We have predicted analytically 
the preferred ripple wavelength, the wave speed and the threshold saltation 
flux required for
ripples to form.
Our numerical simulations show the development of realistic ripple profiles 
from initial surface roughness via growth and ripple merger.

We thank Neal Hurlburt and Alastair Rucklidge for the use of
their compact finite differences code.
The work of RBH was supported by King's College, Cambridge. The work of AM 
was supported by an EPSRC Visiting Fellowship at Oxford Physics.

\vfil\eject
\onecolumn

\end{document}